\documentclass{PoS}

\newcommand{\be}{\begin{equation}}
\newcommand{\ee}{\end{equation}}
\newcommand{\bary}{\begin{eqnarray}}
\newcommand{\eary}{\end{eqnarray}}

\title{Flaring activity of Mrk 421 in 2012 and 2013: orphan flare and multiwavelength analysis}

\ShortTitle{Mrk421}

\author{\speaker{Nissim Fraija}\thanks{Luc Binette postdoctoral scholarship.}\\
        Instituto de Astronomía, Universidad Nacional Aut{\'o}noma de M{\'e}xico\\
        E-mail: \email{nifraija@astro.unam.mx}}

\author{Jos\'e Ignacio Cabrera\\
        Facultad de Ciencias, Universidad Nacional Aut{\'o}noma de M{\'e}xico\\
        E-mail: \email{jcabrera@ciencias.unam.mx}}

\author{Erika Ben{\'i}tez\\
        Instituto de Astronomía, Universidad Nacional Aut{\'o}noma de M{\'e}xico\\
        E-mail: \email{erika@astro.unam.mx}}
        
\author{David Hiriart \\
        Instituto de Astronomía, Universidad Nacional Aut{\'o}noma de M{\'e}xico\\
        E-mail: \email{hiriat@astro.unam.mx}}

\abstract{The well-known blazar Mrk 421 is one of the most active and bright extragalactic sources in the X-ray/$\gamma$-ray sky. In 2012 and  2013, this object displayed two flares or high activity states detected by Fermi-LAT. The first one started in 2012 July 16 (MJD 56124)
and the second one in 2013 April 9 (MJD 56391). The multiwavelength data analysis shows that the $\gamma$-ray flare observed in 2012 was not detected in the hard-X ray bands. This result is usually interpreted as an "orphan" flare. In 2013, the analysis of the
multiwavelength light curves shows that there are two very bright states detected in the optical R-band. The first one in 2013 April 9 (R =11.74 $\pm$ 0.04) and the second one in May 12 (R =11.62 $\pm$ 0.04). Also, high activity states were detected in the soft and hard X-rays. A discrete correlation function analysis of this last flare shows a strong correlation between the GeV $\gamma$-rays and the optical/hard-X ray emission. These results are discussed in terms of the more adequate standard scenarios that could explain the multiwavelength variations displayed by this blazar.}
\FullConference{Swift: 10 Years of Discovery,\\
		2-5 December 2014\\
		La Sapienza University, Rome, Italy }

\begin{document}

\section{Introduction}
At a distance of 134.1 Mpc, the BL Lac object Mrk 421 (z=0.03) \cite{2005ApJ...635..173S} is one of the closest sources in the extragalactic sky. In the MeV - TeV  energy range, simultaneous observations have been carried out with the Fermi-LAT satellite \footnote{http://fermi.gsf.nasa.gov/ssc/data/}. This blazar was also studied using different telescopes based on Imaging Atmospheric Cherenkov Techniques (e.g. VERITAS, MAGIC, H.E.S.S.) and air shower arrays (e.g. ARGO-YBJ, HAWC).  In X-rays, this object has been observed with Swift-XRT and Swift-BAT instruments for almost 10 years  \footnote{http://swift.gsfc.nasa.gov/cgi-bin/sdc/ql?}. In the optical R-band, this source has been monitored since 2008 at San Pedro M{\'a}rtir Observatory as part of the GASP-WEBT program \footnote{http://www.oato.inaf.it/blazars/webt/}  \cite{2008A&A...481L..79V}. On the other hand,  VERITAS observatory and the Whipple 10m Cherenkov telescope observed  the flaring activity in TeV $\gamma$-rays and studied the possible correlations with  X-rays, optical and radio wavelengths between 2006 January and 2008 June  \cite{2011ApJ...738...25A}.  Acciari et al. (2011) reported no significant flux correlations between the TeV $\gamma$-rays and the optical/radio bands. However, an enhanced active phase in the X-ray and in TeV $\gamma$-ray bands was observed. During this active phase, they found strong X-ray activity with no increased TeV emission. Later, a TeV $\gamma$-ray activity lasting two days was detected without  activity in X-rays. Therefore, the later was associated with two "orphan" flares. At the end of this active state, the source showed a significant correlation between the X-ray and TeV bands.  In general, no correlation in TeV has been found with optical and/or radio fluxes \cite{2011ApJ...738...25A}. However, correlations among X-ray and TeV bands have been reported in \cite{1995ApJ..449..99, 2008ApJ..677..906,2009ApJ..695..596}. 
Although most of these correlations are interpreted through the standard one-zone SSC model (synchrotron self-Compton; \cite{2008ApJ...686..181F}), other correlations suggest serious deviations from this leptonic model.

The Spectral Energy Distribution (SED) of Mrk 421 presents a double-humped shape; a low energy hump at energies $\simeq$ 1 keV  and the second hump at hundreds of GeV. Abdo et al. (2011) found that both leptonic and hadronic scenarios  are able to fit the Mrk 421 SED reasonably well, implying comparable jet powers but very different characteristics for the blazar emitting region. In the leptonic scenario, a one-zone SSC with three accelerated electron populations (through diffusive relativistic shocks with a randomly oriented magnetic field)  has been used \cite{2011ApJ...736..131A}.  In the hadronic scenario \cite{2014arXiv1411.7354F}, the peak at low energies is explained by electron synchrotron radiation whereas the high-energy peak is explained invoking the Synchrotron Proton Blazar (SPB) model \cite{2001APh....15..121M,2003APh....18..593M}. In this work, we show the multiwavelength observations carried out on Mrk 421 during the active states displayed in 2012 and 2013 in the GeV energy range. A brief discussion related to the theoretical interpretations of both flares will be given.   
\section{Multiwalength Light Curves}
Multiwavelength light curves of the 2012 and 2013 flares of Mrk 421 are shown in Figure 1.  The GeV $\gamma$- ray data shown corresponds to the 200 MeV to 300 GeV band. Details on the reduction procedure applied to these data-set can be found in \cite{2013MNRAS.434L...6C}. The X-ray data were obtained with both Swift-BAT (15 - 50 keV) and Swift-XRT (0.2 - 10 keV) instruments. The optical R-band observations were carried out with the 0.84 m f/15 Ritchey-Chretien telescope and the instrument POLIMA \footnote{A detailed description of our photopolarimetric monitoring program on TeV blazars can be found in http://www.astrossp.unam.mx/blazars}.  It is worth noting that the optical R-band magnitudes are not corrected for the contribution of the host galaxy of Mrk 421.  Additionally, and only for comparison, we have included a few optical R-band data collected by the American Association of Variable Star Observers (AAVSO) \footnote{http://www.aavso.org/observing-campaigns}.
\subsection{2012 Flare}
Mrk 421 was detected very active in the GeV energy range  in July 16 with a daily flux of (1.4$\pm$0.2)~$\times10^{-6}$ ph cm$^{-2}$ s$^{-1}$, see ref. \cite{2012ATel.4261....1D}. The source continued to be detected between July 17 and 21, with a daily flux between (0.4$\pm$0.1)$\times~10^{-6}$ ph cm$^{-2}$ s$^{-1}$ and (0.9$\pm$0.2)~$\times10^{-6}$ ph cm$^{-2}$ s$^{-1}$. 
In fig. 1 we highlight the active state using a red vertical bar. It is clear that $\gamma$-ray flare was detected without activity in the hard X-ray band. Unfortunately,  data were collected neither in the soft-X rays nor the optical R-band. 
\subsection{2013 Flare}
Fermi-LAT reported high activity of Mrk 421 in 2013 April 9 to 12, with a daily flux between (0.4$\pm$0.1)$\times~10^{-6}$ ph cm$^{-2}$ s$^{-1}$ and (0.8$\pm$0.2)$\times~10^{-6}$ ph cm$^{-2}$ s$^{-1}$, see ref. \cite{2013ATel.4977....1P}. In fig. 1, the states of high activity  have been divided in two vertical color bands, purple and green. The purple one marks the high activity observed in GeV $\gamma$-rays, X-rays (XRT and BAT) and optical bright R-band (April 9, R =11.74 $\pm$ 0.04) . The green color vertical band marks the 
second bright optical R-band point (May 12, R=11.62 $\pm$ 0.04).  The second optical bright point seems to be anti-correlated with the higher energy bands. 

\section{Discussion}

From the multiwavelength light curves it is clear that the  $\gamma$-ray flare observed in 2012 was detected  without any strong activity in the hard-X rays.   In 2012 July 16, a TeV $\gamma$-ray flare without any increased activity in other wavelengths was reported  \cite{2012ATel.4272....1B}.  The so-called "orphan'' flares have been previously observed in Mrk 421 \cite{2005ApJ...630..130B,2011ApJ...738...25A}, and  also in the blazar  1ES 1959+650  \cite{2003ApJ...583L...9H,2004ApJ...601..151K}.  In general, most of the flaring activity in this source occurs quasi-simultaneously with $\gamma$-ray and X-ray emission. Therefore,  this atypical flaring event observed in the TeV/GeV $\gamma$-rays along with the absence of activity in the X-rays is very difficult to reconcile with the SSC model.  Orphan flares have been usually explained as due to neutral pion decays from proton-photon interactions \cite{2005ApJ...621..176B,2013PhRvD..87j3015S, 2015arXiv150104165F}. It is worth mentioning that a radio flare with a delay of $\sim$ 60 days was detected by the Owens Valley Radio Observatory (OVRO) 40-m Telescope \cite{2012ATel.4451....1H, 2015arXiv150107407H}.\\

Based on the multiwavelength light curves shown in fig. 1,  and the emission in TeV and X-ray reported by \cite{2013ATel.4976....1C, 2014A&A...570A..77P}, it is clear that Mrk 421 flared in TeV/GeV $\gamma$-rays, in X-rays (BAT and XRT) and in optical R-band in 2013 April 9 - 12. The results obtained from the discrete correlation function (DCF) calculated using all the light curves in this work are presented in Figure 2. Left panel shows the correlation between GeV $\gamma$-rays  and hard-X rays, and the right panel shows the correlation between GeV $\gamma$-rays and the optical R-band. In both panels the DCF show that there are no lags between the GeV $\gamma$-rays and hard-X rays, and also with the optical R-band.  Therefore, the multiwavelength emission seems to take place simultaneously in all bands,  which favors a one-zone SSC model.   In the framework of this SSC model, the electrons within the emitting region are moving at ultra-relativistic velocities in a collimated jet. The Fermi-accelerated electrons injected into the emitting region are confined by a magnetic field. Then, photons are radiated via synchrotron emission and up-scattered to higher energies. The low energy emission, from radio to X-rays, is produced by synchrotron radiation. The high energy emission (MeV - TeV) $\gamma$-rays, is due to Compton scattering.  This leptonic model depends basically on the bulk Lorentz factor, the size of emitting region, the electron number density and the strength of the magnetic field.  It is possible to find a set of parameters that can describe the states of low or high activity \cite{2011ApJ...738...25A}. \\

It is worth noting that  the maximum brightness in the optical R-band  observed in May 12 is anti-correlated with the other bands.  This result poses a challenge for the theoretical models proposed for this blazar.  In a forthcoming paper, we will present a more detailed analysis of these active states.

\begin{figure*}
\centering
\includegraphics[width=0.95\textwidth]{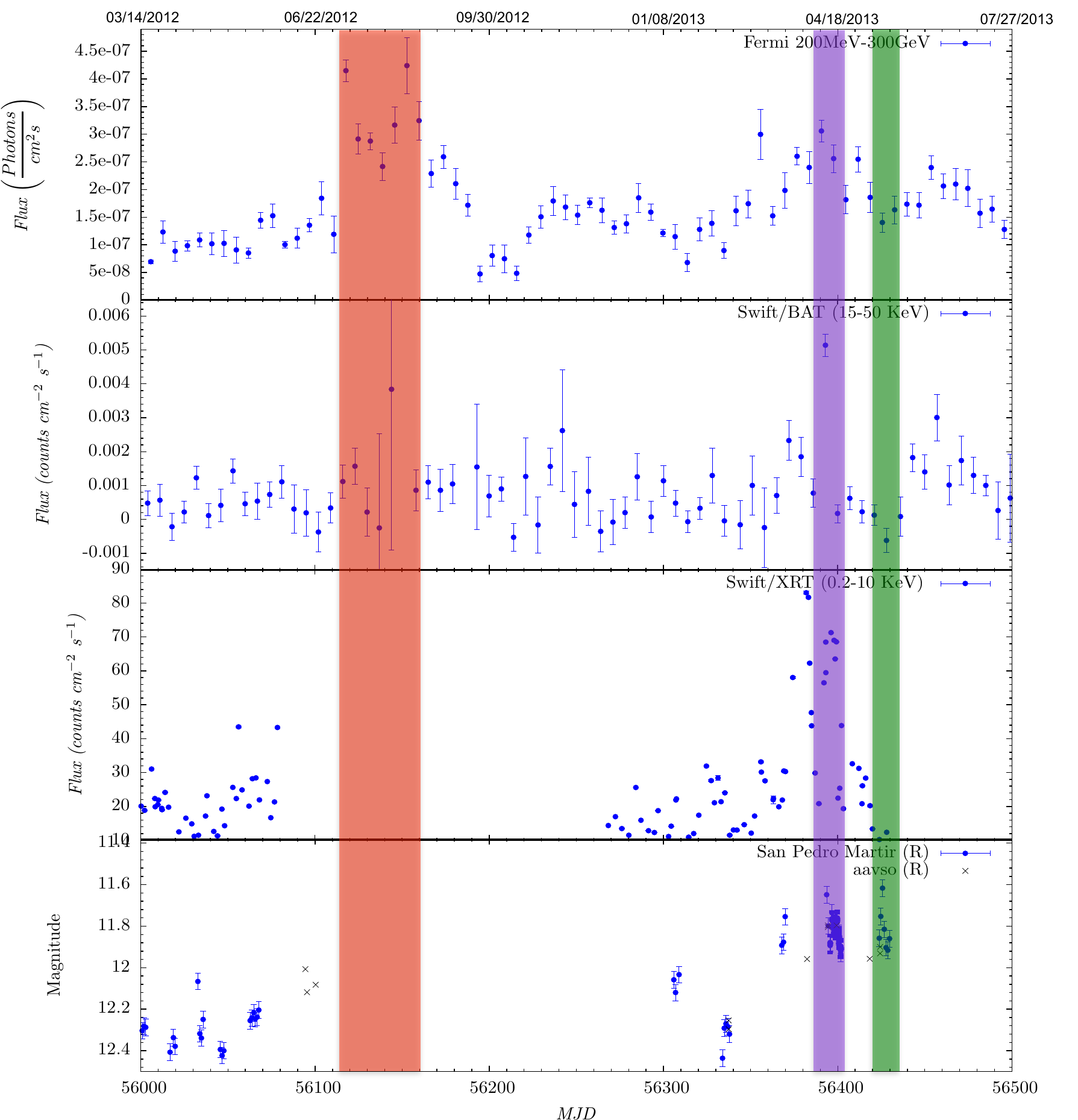}
\caption{Mrk 421 light curves obtained with Fermi,  Swift and  San Pedro M{\'a}rtir data in 2012-2013. From top to bottom: GeV $\gamma$-rays, hard-X rays, soft-X rays and the optical R-band.  \label{light curve}}
\end{figure*}
\begin{figure*}
\centering
\includegraphics[width=\textwidth]{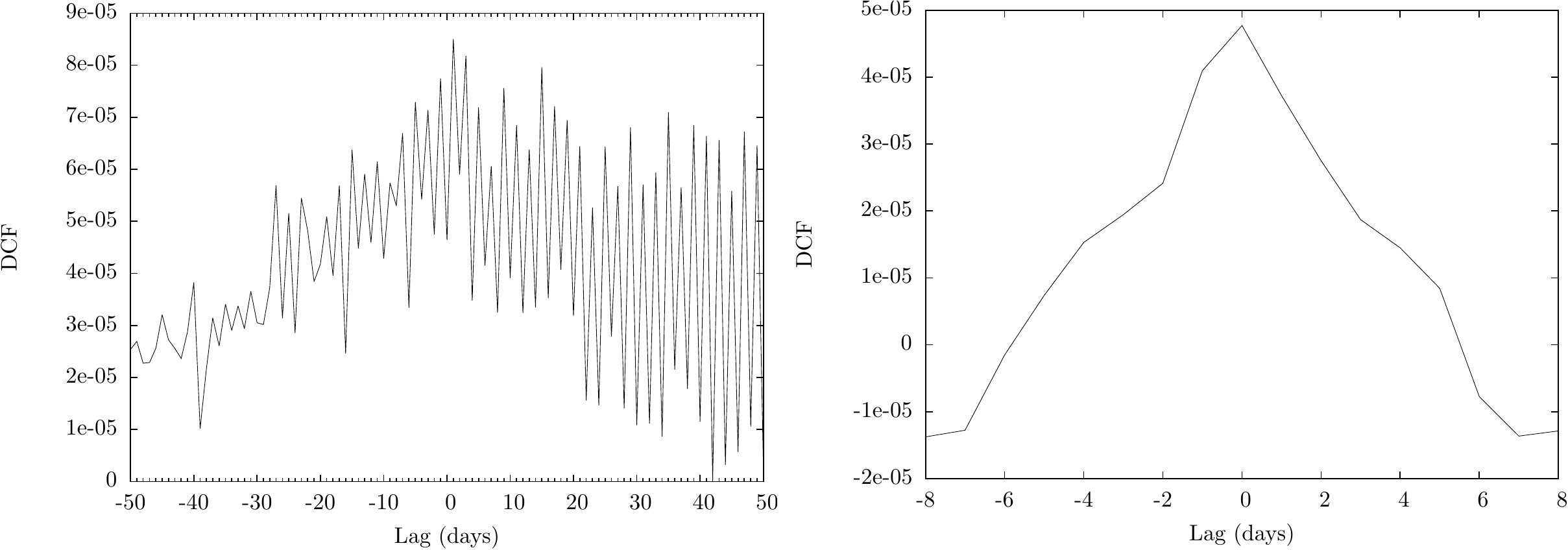}
\caption{Left: DCF  between the GeV $\gamma$-rays and X-rays. Right: DCF between the GeV $\gamma$-rays and the  optical R-Band.    \label{activity:2013}}
\end{figure*}

\end{document}